# Cryptocurrency Address Clustering and Labeling

Mengjiao Wang, Hikaru Ichijo, Bob Xiao

Binance.com

## 1. Introduction

Anonymity is one of the most important qualities of blockchain technology. For example, one can simply create a bitcoin address to send and receive funds without providing KYC to any authority. In general, the real identity behind cryptocurrency addresses is not known, however, some addresses can be clustered according to their ownership by analyzing behavioral patterns, allowing those with known attribution to be assigned labels. These labels may be further used for legal and compliance purposes to assist in law enforcement investigations.

In this document, we discuss our methodology behind assigning attribution labels to cryptocurrency addresses. According to CoinMarketCap[1], as of October 2019, there are more than three thousands cryptocurrencies in existence. Generally, we separate cryptocurrencies into two groups, each using a different method to assign address labels. The first group includes cryptocurrencies with an Unspent Transaction Output (UTXO), such as Bitcoin and Litecoin. Common spending and one-time change addresses are commonly used methods to cluster addresses. The second group focuses on account-based cryptocurrencies, such as Ethereum (including ERC20 tokens) and EOS, and we identify exchange addresses through analyzing funding patterns. Our initial address labels come primarily from two data sources: publicly available crowdsourcing websites such as the Bitcoin Abuse Database[2] and Etherscan[3]; as well as data annotated by Binance customers.

## 2. Related Works

Since the birth of Bitcoin, both anonymization and de-anonymization techniques have been studied within the crypto sphere. One such example is the use of common spending and one-time change addresses [2][3] to cluster addresses for UTXO-based cryptocurrencies, such as Bitcoin.

For account-based cryptocurrencies, like Ethereum, addresses are typically reused and do not change as frequently as UTXO-based cryptocurrencies. Although many address labels are published through official and unofficial channels (e.g. Etherscan for Ethereum and Bithomp for XRP), the vast majority of accounts still lack attribution. However, by using available labels as seeds, we may expand address labels to include deposit addresses belonging to cryptocurrency exchanges, dependent on their wallet architecture [5].

---

[1] CoinMarketCap https://coinmarketcap.com/
[2] Bitcoin Abuse Database https://www.bitcoinabuse.com/
[3] Etherscan https://etherscan.io/

## 3. Address Clustering

### a. UTXO-based Cryptocurrencies

#### i. Common Spending

Common spending is one of the most commonly used methods for clustering Bitcoin addresses. For example, in Figure 1, the three input addresses for the transaction can be attributed to the same owner. This is because, in order to initiate the transaction, the sender must know the private keys for all input addresses at the same time.

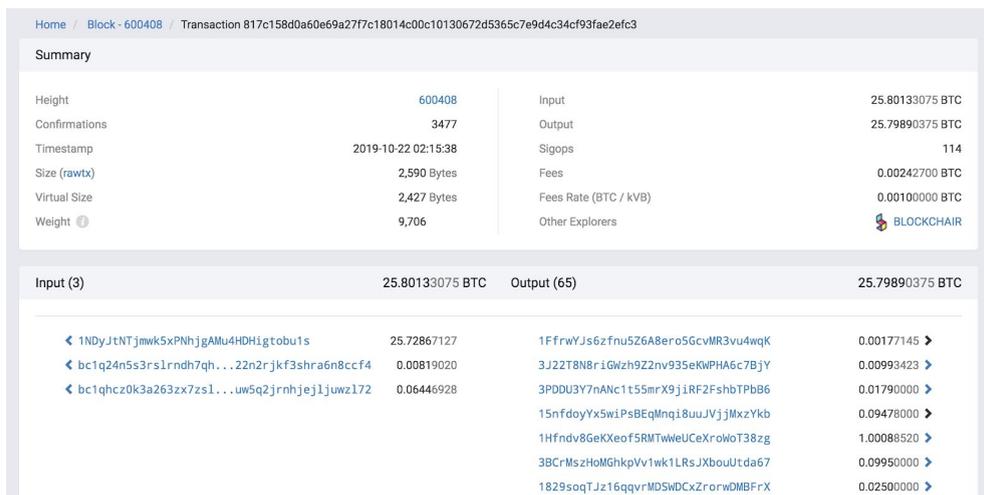

**Figure 1: Screenshot of a Bitcoin Transaction**

We can create clusters by first building a graph with only the input addresses and transactions themselves (Figure 2), which then allows us to identify new relationships within the graph. Each linked component can be treated as a part of the group.

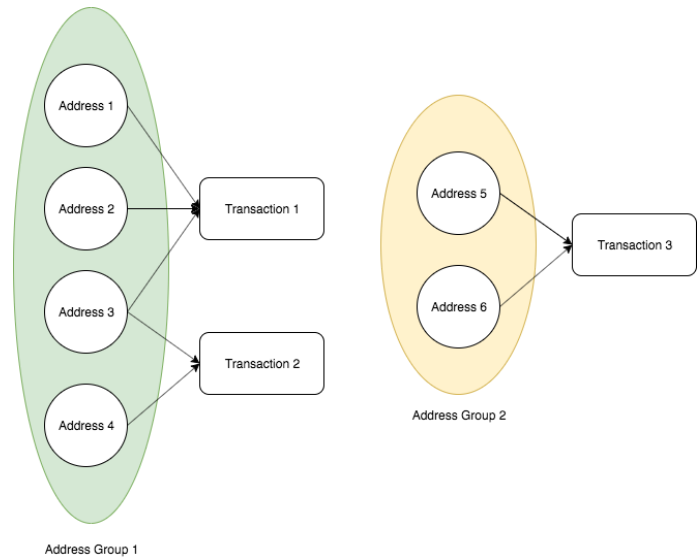

**Figure 2: Clustering via Transaction Graph**

Common spending assumes that each Bitcoin address is controlled by a single entity in the real world. Although not necessarily true for multi-signature addresses, the multi-sig technique does not change the nature of address clustering, and there remains an implied association between the entities.

### ii.  One-time Change Inference

The change resulting from a Bitcoin transaction is returned to the sender using a new address located on the output side. Therefore, change address analysis offers another way to expand upon clusters. However, as change addresses are not explicitly marked in a Bitcoin transaction, and they cannot always be inferred correctly, we must develop a number of patterns to minimize false-positives.

An address has a high possibility of being a change address if it fits the following patterns:

- there is more than one input address in the transaction;
- one output address is new, but the remainder have been used;
- the transfer amount of the new output address extends greater than 4 digits following the decimal;
- the input and output share no common addresses;

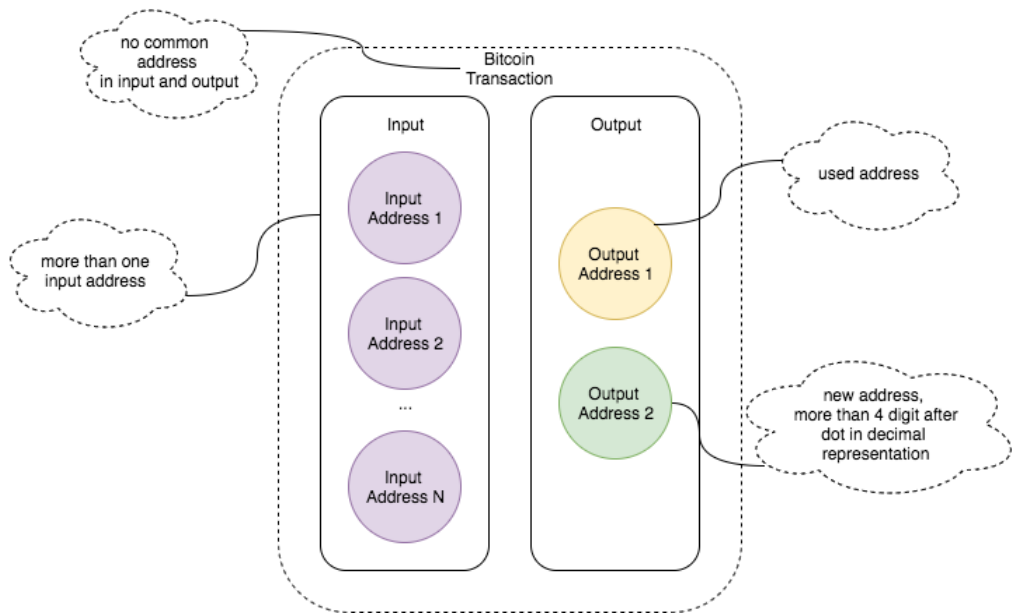

**Figure 3: One-time Change Patterns**

### b. Account-based Cryptocurrencies

Unlike UTXO-based cryptocurrencies, change is not needed as senders may always send exact quantities of assets to recipients. Most cryptocurrency exchanges generate dedicated blockchain addresses for customers to process deposits. Assets received in the dedicated address will later be automatically transferred to a hot wallet. As exchange hot wallets are usually public and labeled through crowdsourcing, we can infer customer addresses using this gathering pattern.

### i. Fund Gathering Patterns

Fund gathering patterns are used to cluster addresses that are created by cryptocurrency exchanges for customer deposits. If the addresses are deemed to belong to the same exchange, they are then grouped together.

An example of fund gathering is as follows:

**Figure 4: Ethereum is Swept to a Hot Wallet**

A customer had 174.65893626 Ethereum and wanted to deposit it to their Binance account. To accomplish this, the customer simply transferred Ethereum to the deposit address for their account (provided and controlled by Binance). Some time after the transaction was confirmed, Binance automatically migrated the assets from that deposit address on-chain to the exchange's hot wallet address.

This gathering pattern is summarized in Figure 5, showing that we may group customer deposit addresses, as well as exchange hot and cold wallets, into a single cluster.

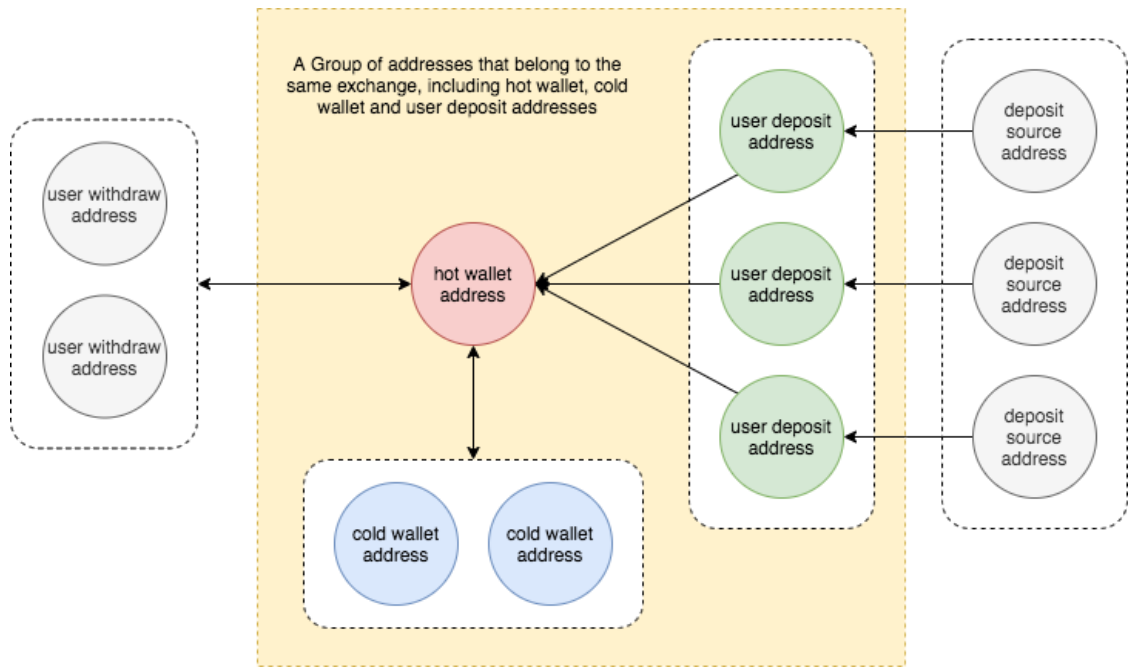

**Figure 5: Clustering Exchange Addresses using Gathering Patterns**

Hot and cold wallets for most exchanges can be obtained from public sources (e.g. Etherscan for Ethereum). Once a hot wallet is correctly labeled, we may then identify customer deposit addresses using the hot wallet. Usually, a customer's deposit address transfers assets only to the hot wallet address, but may receive assets from any source.

### 4. Experiments

We evaluated our methodology using the public transactional data of Bitcoin and Ethereum networks before November 2019. Detailed results are as follows:

**a. Bitcoin**

Executing our technique on the Bitcoin network, we were able to assign labels for 115 million addresses. The top ten clusters with the most addresses are shown below:

| category | name | number of addresses |
|---|---|---|

| category | peer_name | number of addresses |
|---|---|---|
| exchange | coinbase.com | 23,189,630 |
| merchant service | gocoin.com | 6,960,885 |
| p2p exchange | localbitcoins.com | 5,793,274 |
| merchant service | coinpayments.net | 5,080,340 |
| merchant service | bitpay.com | 4,343,764 |
| exchange | binance.com | 3,663,255 |
| hosted wallet | xapo.com | 2,171,501 |
| exchange | coins.co.th | 2,087,623 |
| exchange | cubits.com | 2,043,583 |
| exchange | bittrex.com | 1,604,589 |

### b. Ethereum

Using our methodology, we were able to label more than 7 million addresses from a seed set of only 515 exchange addresses. The top ten exchanges with most customer deposit addresses are below:

| category | peer_name | number of addresses |
|---|---|---|
| exchange | binance.com | 2,175,948 |
| exchange | bittrex.com | 804,789 |
| exchange | kucoin.com | 600,525 |
| exchange | bitfinex.com | 352,903 |
| exchange | kraken.com | 339,253 |
| exchange | huobi.com | 330,841 |
| exchange | poloniex.com | 272,129 |
| exchange | okex.com | 265,791 |
| exchange | upbit.com | 234,105 |
| exchange | hitbtc.com | 224,654 |

## 5. Conclusion

- We demonstrated address clustering techniques for both UTXO-based and account-based cryptocurrencies;

- Our methodology was implemented and evaluated with the Bitcoin network and Ethereum network. In total, 115 million Bitcoin addresses and 7 million Ethereum addresses were assigned a label.